# Seeing many-body effects in single- and few-layer graphene: Observation of two-dimensional saddle-point excitons

Kin Fai Mak[1], Jie Shan[2], and Tony F. Heinz[1*]

[1]Departments of Physics and Electrical Engineering, Columbia University, 538 West 120th St., New York, NY 10027, USA

[2]Department of Physics, Case Western Reserve University, 10900 Euclid Avenue, Cleveland, OH 44106, USA

Significant excitonic effects were observed in graphene by measuring its optical conductivity in a broad spectral range including the two-dimensional π-band saddle-point singularities in the electronic structure. The strong electron-hole interactions manifest themselves in an asymmetric resonance peaked at 4.62 eV, which is red-shifted by nearly 600 meV from the value predicted by *ab-initio* GW calculations for the band-to-band transitions. The observed excitonic resonance is explained within a phenomenological model as a Fano interference of a strongly coupled excitonic state and a band continuum. Our experiment also showed a weak dependence of the excitonic resonance in few-layer graphene on layer thickness. This result reflects the effective cancellation of the increasingly screened repulsive electron-electron (*e-e*) and attractive electron-hole (*e-h*) interactions.

A fundamental issue in understanding the unique properties of electrons in graphene, a single layer of carbon atoms, is the role of many-body interactions in this system. Many-body interactions are generally more significant in two-dimensional (2D) materials than in their bulk counterparts. This reflects both the intrinsic enhancement of the importance of Coulomb interactions in 2D materials, as well as their reduced screening. Indeed, the behavior of low-energy quasiparticles in graphene is significantly altered by *electron-electron* (*e-e*) interactions, as manifested by recent reports of broken symmetry states in the quantum Hall regime [1], of the fractional quantum Hall effect [2], of a renormalization of the Fermi velocity [3], and of plasmarons [4]. The role of *electron-hole* (*e-h*) interactions, as are relevant for the optical response, has remained somewhat paradoxical. On the one hand, theoretical studies have predicted strong excitonic corrections to the optical response [5-9], as well as BCS superfluidity at high quasiparticle density [10] and magnetoexcitonic superfluidity under high magnetic fields [11]; on the other hand, the vast preponderance of experimental data, including the optical conductivity in the infrared, can be understood within an independent-particle description of carriers in linearly dispersing bands [12-16].

In this Letter, we present direct experimental evidence of large excitonic corrections to the optical response of graphene. We do so by measuring graphene's optical conductivity $\sigma(E)$ over a wide range of photon energies $E = 0.2 - 5.3$ eV. Above the infrared spectral range, we observe that $\sigma(E)$ starts to deviate significantly from the universal value of $\pi e^2/2h$ predicted within an independent-particle model [17] and demonstrated experimentally [15, 16]. A pronounced resonance feature arises in the ultraviolet from transitions near the *M*-point, a saddle-point singularity, of the Brillouin zone of graphene [18]. The feature is well known in graphite [19] and has recently been observed in graphene [20]. While the enhanced absorption associated with this feature is expected within an independent-particle picture, the marked asymmetry of the line shape cannot be understood [21]. Further, the measured position of the peak, when compared with theory, also indicates a large red shift (~ 600 meV) [7]. We attribute these features to strong *e-h* interactions of the quasiparticles near the 2D saddle-point singularity. The distinctive asymmetric line shape can be viewed as an interference effect in the classic theory of Fano [22, 23] for a discrete state residing within a continuum. The detailed spectral dependence of this excitonic resonance is found to be in excellent agreement with the result of recent first-principles calculations [7]. In addition, only a very slight variation of excitonic resonance with layer thickness was observed in few-layer graphene (FLG), despite the increased screening of the Coulomb interactions for thicker samples. This weak dependence on screening reflects the effective cancellation of the repulsive *e-e* interactions, which shift the transition upwards in energy, and the attractive *e-h* interactions, which shift it downwards [7].

In our measurements, we made use of graphene samples prepared by mechanical exfoliation of kish graphite (Toshiba) on $SiO_2$ substrates (Chemglass). The procedure for sample preparation and characterization is described in [16, 24, 25]. The optical response was determined using different sources for different wavelength regimes. For the mid-to-near IR range (0.2 – 1.2 eV) radiation from the National Synchrotron Light Source at Brookhaven National Laboratory was employed together with a Fourier-transform infrared microscope [16, 24]. The visible (1.5 – 3.0 eV) and UV (3.0 – 5.3 eV) ranges were examined in our laboratory with, respectively, a quartz tungsten halogen (QTH) and a deuterium source, coupled with a confocal microscope. The reflected beam was analyzed with a grating spectrometer and a UV-enhanced CCD camera.

We obtain the optical conductivity of graphene $\sigma(E)$ directly from the reflectance contrast $\Delta R/R$, the normalized change in reflectance of the bare substrate induced by the presence of the sample. For a sufficiently thin sample supported on a transparent substrate with refractive index $n_s$, $\sigma(E)$ is given by $\sigma(E) = \frac{c}{4\pi}\frac{n_s^2-1}{4}\frac{\Delta R}{R}$ [16]. Within the spectral range of our measurements the SiO$_2$ substrate has negligible absorption and the dispersion of $n_s$ is well established [26]. The thin sample approximation also holds reasonably well for graphene thickness up to 5 layers [24]. Thus, we were able to obtain $\sigma(E)$ (or, equivalently, the absorption spectrum) of graphene samples directly.

The measured optical conductivity spectrum $\sigma(E)$ for a graphene monolayer (Fig. 1) displays several noteworthy features. (i) In the near-IR spectral range of 0.5 – 1.5 eV, $\sigma(E)$ is described well by the universal value of $\sigma = \pi e^2/2h$, as derived previously within the independent-particle theory [17] and confirmed experimentally [15, 16]. (ii) In the visible spectral range, $\sigma(E)$ rises smoothly and steadily, increasing by ~ 80% at 3.0 eV. (iii) In the UV range, $\sigma(E)$ displays a pronounced peak at $E_{exp}$ = 4.62 eV. In the independent-particle description, this feature arises from interband transitions in graphene from the bonding to the antibonding $\pi$-states near the saddle-point singularity at the $M$-point of the Brillouin zone [18]. GW calculations, which are known to be accurate in describing the quasiparticle bands of graphene (but ignore the $e$-$h$ interactions), predict a band-to-band transition energy of $E_{GW}$ = 5.2 eV at the $M$-point [Fig. 2(b)] [7, 27]. The observed peak energy $E_{exp}$ is red-shifted from $E_{GW}$ by almost 600 meV, over 10% of the saddle-point energy. (iv) The observed resonance feature has an asymmetric line shape, with higher absorption on the low photon-energy side. Such an asymmetry is also in disagreement with the single-particle description. In this description, the optical transition matrix elements depend only weakly on the energy $E$ in the vicinity of the 2D saddle-point singularity. The spectral dependence of the optical conductivity is then governed by the joint density of states (JDOS), -$log$ $|1-E/E_0|$ [21], which is symmetric in energy about the saddle-point energy $E_0$.

A natural way to account for the observed discrepancies from the independent-particle description of the optical response of graphene is to include $e$-$h$ interactions. Effects of the $e$-$h$ interaction near a saddle-point singularity have been extensively investigated theoretically [28]. Such interactions give rise to significant changes in the line shape near the saddle point. This redistribution of oscillator strength can be understood in terms of the development of discrete excitonic states lying below the saddle-point singularity from the attractive $e$-$h$ interactions. These excitonic states do not, however, lie below a true gap. They consequently couple strongly with the existing continuum of electronic states, leading to asymmetric resonance features with enhanced absorption on the low-energy side [23, 28, 29]. Philips [23] has interpreted the overall absorption line shape in terms of just such a Fano interference [22, 23] effect. We extend this phenomenological approach to graphene by including a single dominant excitonic state. We express the resultant optical conductivity $\sigma(E)$ as:

$$\frac{\sigma(E)}{\sigma_{cont}(E)} = \frac{(q+\varepsilon)^2}{1+\varepsilon^2}. \qquad (1)$$

Here $\sigma_{cont}(E)$ denotes the optical conductivity arising from the unperturbed band-to-band transitions (*i.e.*, without *e-h* interactions); $\varepsilon = (E - E_{res})/(\Gamma/2)$ is the normalized energy relative to the resonance energy $E_{res}$ of the perturbed exciton (*i.e.*, with coupling to the continuum). The parameter $q^2$ defines the ratio of the strength of the excitonic transition to the unperturbed band transitions, while the sign of $q$ determines the asymmetry of the line shape.

The behavior predicted by the phenomenological Fano model is shown in Fig. 2(a). We model the unperturbed band transitions by the JDOS near the saddle-point singularity, $\sigma_{cont}(E) \sim -log\ |1-E/E_0|$, broadened by convolution with a Lorentzian of width $0.01E_0$ to account for broadening from carrier relaxation. For the appropriate Fano parameters ($q = -1$, $E_{res} = 0.95E_0$, $\Gamma = 0.01E_0$), we obtain a spectrum with a peak red-shifted from $E_0$ and an asymmetric line shape with enhanced optical conductivity on the low-energy side. Despite its simplicity, we see that this model captures the main features of the experimental optical conductivity spectrum $\sigma(E)$. A more quantitative analysis can be carried out using the results of published GW calculations for the continuum background [7], which provide a better description of the response well away from the singularity. We consequently take $\sigma_{cont}(E) = \sigma_{GW}(E)$, where we have convoluted the theoretical results with broadening of 250 meV to reflect the experimentally observed width. The best fit to Eq. 1 then yields $q = -1$, $E_{res} = 5.02$ eV, and $\Gamma = 780$ meV. The exciton resonance energy is seen to be red-shifted by 180 meV from the GW peak at $E_{GW} = 5.20$ eV. The large value of the width parameter $\Gamma$ corresponds to an exciton lifetime of only $\sim 0.5$ fs, indicating a very high rate of autoionization of the excitonic state into the continuum. We also compare the measured $\sigma(E)$ to the recent result of full *ab-initio* calculations in which excitonic effects were taken into account using the GW-Bethe-Salpeter approach (GWBS) [7] [dashed line of Fig. 2(b), broadened by 250 meV]. Both the GWBS calculations and the phenomenological model provide excellent agreement with the available experimental data.

Although the effects of *e-h* interactions near a saddle point are known to be significant in insulators and semiconductors [28], the observed excitonic effects in graphene must be considered in terms of the dielectric response of a *semimetal*. In a highly conductive system like graphene, one might expect effective screening of the Coulomb potential and, consequently, weak *e-h* interactions. Indeed, theoretical investigations have shown that screening of charges in the 2D graphene system arises from a combination of "metallic" screening by intraband transitions and "insulating" screening from interband transitions [30]. The former is similar to screening in other 2D electron systems, but the relatively low DOS near the Dirac point, a consequence of the linear dispersion relation, reduces this contribution compared to that which would be present in a normal metal. In addition, screening in graphene is diminished by the fact that the charges can act upon one another with electric field lines that extend into the surrounding media, which usually exhibit a weaker dielectric response. As a consequence, the Coulomb interaction remains quite effective in graphene and accounts for the pronounced excitonic corrections that we observe experimentally.

How does the excitonic resonance vary with layer thickness $N$ in FLG? While the band structure near the M-point is modified by interlayer interactions [27, 31], because of the symmetry properties of the electronic states, only optical transitions with very similar energies are dipole allowed. Thus, the band structure changes are not expected to lead to a significant shift in the optical resonance, as shown below [32]. On the other hand, as the thickness of the sample increases towards the bulk limit the screening of the *e-h* interactions is expected to become considerably more effective because of the reduced influence of the surrounding environment and the increased density of states at the Fermi energy [31]. In fact, however, only very slight changes in the excitonic resonance were observed in our experiment for $N = 1 - 5$ layers [Fig. 3(a)]. The optical conductivity increases almost linearly with layer thickness, except a small red shift of the resonance peak $E_{exp}$ [Fig. 3(b)].

To understand this unexpected behavior of the excitonic resonance in FLG, let us examine the peak positions calculated within several *ab-initio* approaches to the electronic structure: $E_{LDA}$ from the density-functional theory (DFT) in the local-density approximation [32], $E_{GW}$ from the GW corrections to DFT [7], and $E_{GWBS}$ from the GW-Bethe-Salpeter treatment [7]. $E_{LDA}$ underestimates the peak energy $E_{exp}$, but like the experimental data depends very weakly on $N$, reflecting the omission of both long-range *e-e* correlations and *e-h* interactions [27]. It also shows that band structure changes of the material, as a result of interlayer interactions, do not lead to a significant change in the resonance energy. On the other hand, $E_{GW}$ overestimates the peak energy and shifts to lower energies with increasing $N$. This behavior arises from the inclusion of the *e-e* repulsion (but not *e-h* attraction), which is increasingly screened with increasing thickness. Indeed, the peak energy converges rapidly to the bulk (graphite) value for $N > 3$. Now with both *e-e* and *e-h* interactions taken into account within the GW-Bethe-Salpeter approach, $E_{GWBS}$ displays only a slight red shift with increasing $N$, in accord with experiment. Thus, the weak dependence of $E_{exp}$ on $N$ is not the consequence of identical screening of Coloumb interactions for mono- and few-layer graphene. It is rather a result of a significant cancellation of comparable repulsive and attractive Coulomb interactions. The net energy of the excitonic state is thus less sensitive to screening than that of either the quasiparticle band gap or the exciton binding energy taken alone. The phenomenon is analogous to that observed in carbon nanotubes [33-37] and predicted in graphene nanoribbons [5].

In conclusion, our experiment has provided the first direct evidence of strong excitonic effects in the optical response of graphene. In addition to its intrinsic interest, graphene is an excellent model system for the study of 2D saddle-point excitons. Of particular interest would be probing of its behavior at variable doping density.

The authors thank Dr. Mark Hybertsen for fruitful discussions. The authors at Columbia acknowledge support from the U.S. Department of Energy EFRC program (grant DESC00001085) and from the Office of Naval Research under the MURI program; the author at Case Western acknowledges National Science Foundation grant DMR-0907477.

**Figures:**

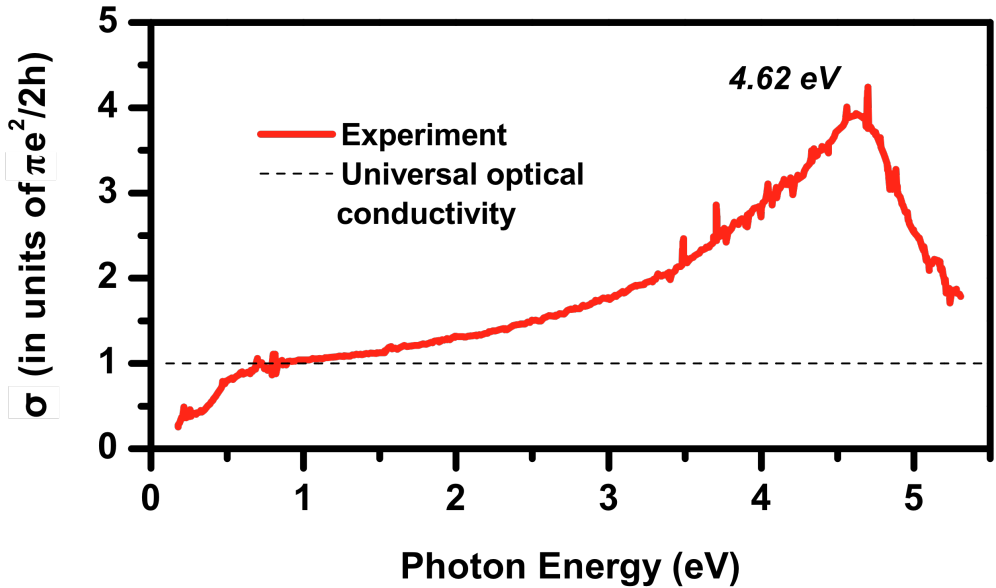

Figure 1: Experimental optical conductivity (solid line) and the universal optical conductivity (dashed line) of monolayer graphene in the spectral range of 0.2 – 5.5 eV. The experimental peak energy is 4.62 eV. Note the deviation of the optical conductivity from the universal value at low energies is attributed to spontaneous doping [16].

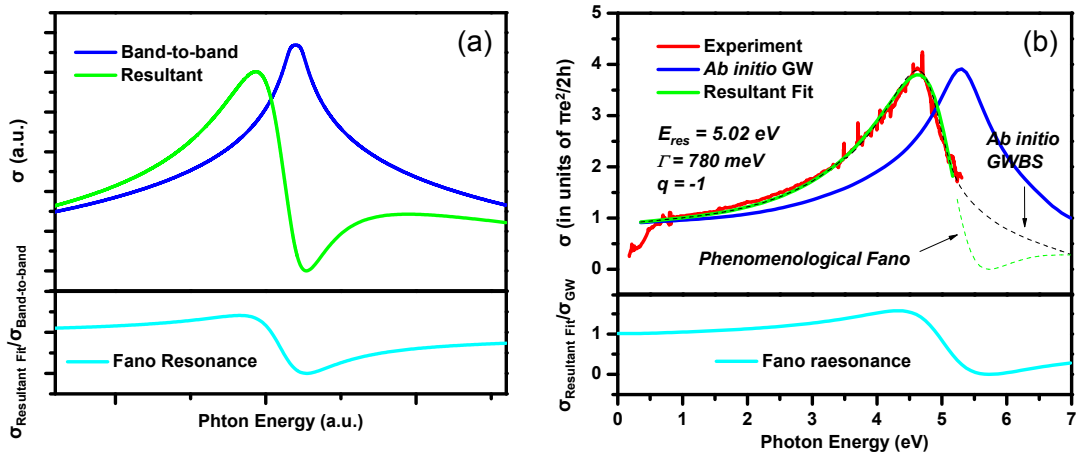

Figure 2: Fano theory of saddle-point excitons in graphene. Upper panels: optical conductivity of unperturbed band-to-band transitions $\sigma_{cont}$ (blue) and the resultant conductivity from a Fano interference of the exciton state and the continuum (green); lower panels: Fano line shape of Eq. 1. (a) Illustration of the model using the JDOS near the 2D saddle point for $\sigma_{cont}$ and the Fano parameters as described in the text. (b) Fit of experiment (red) to the Fano model using the optical conductivity obtained from GW

calculations [7] for $\sigma_{cont}$. The dashed line is the optical conductivity spectrum obtained from the full GW-Bethe-Salpeter calculation [7].

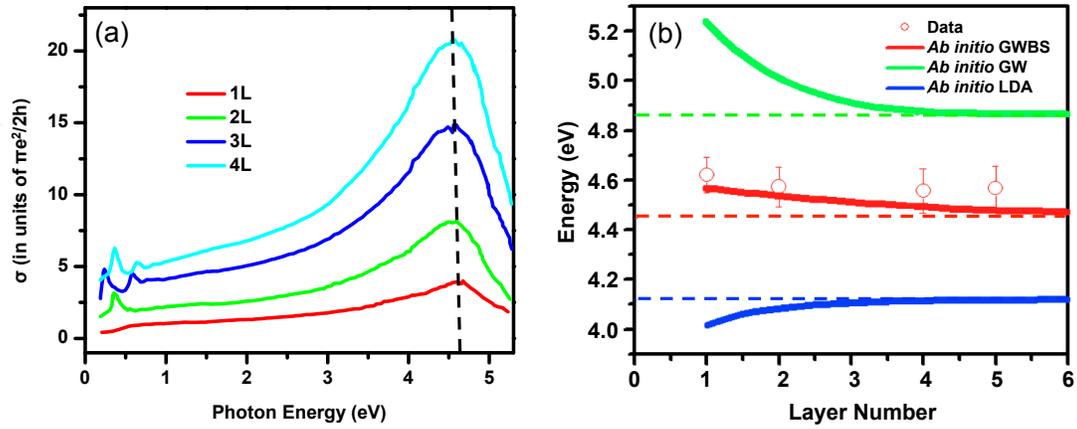

Figure 3: (a) Experimental optical conductivity for single- and few-layer graphene in the spectral range of 0.2 – 5.3 eV. The dashed line is a guide to the eye of the resonance peak. (b) Dependence of the peak energy of the excitonic resonance on layer thickness: experiment (symbols), values calculated from *ab-initio* LDA (blue), GW (green) and GWBS (red). The dashed lines show the calculated value for bulk graphite.